\documentclass[]{article}

\usepackage[]{natbib}

\usepackage{graphicx}
\usepackage{amsmath}
\usepackage[notref,notcite,final]{showkeys}
\usepackage{hyperref}
\usepackage{dsfont}
\usepackage[usenames,dvipsnames]{xcolor}
\usepackage{ulem}

\newcommand\solidline[1][3mm]{\rule[.4ex]{#1}{1pt}}
\newcommand\dashedline{\mbox{%
  \solidline[1mm]\hspace{1mm}\solidline[1mm]}}
\newcommand\refeq[1]{Eq.~(\ref{#1})}
\newcommand\refEq[1]{Equation~(\ref{#1})}
\newcommand\reffig[1]{Figure~\ref{#1}}

\title{Evaluating ensemble forecasts by the Ignorance score -- Correcting the finite-ensemble bias}

\author{Stefan Siegert\footnote{Corresponding author: Stefan Siegert, College of Engineering, Mathematics and Physical Sciences, University of Exeter, Laver Building, North Park Road, Exeter, EX4 4QE, United Kingdom, Email: s.siegert@exeter.ac.uk}, Christopher A. T. Ferro, David B. Stephenson\\{\small University of Exeter, United Kingdom}}

\date{\today}

\begin{document}

\maketitle

\begin{abstract}

This study considers the application of the Ignorance Score (also known as the
Logarithmic Score) in the context of ensemble verification.  In particular, we
consider the case where an ensemble forecast is transformed to a Normal
forecast distribution, and this distribution is evaluated by the Ignorance
Score.  It is shown that the standard Ignorance score is
biased with respect to the ensemble size, such that larger ensembles yield
systematically better expected scores. A new estimator of the
Ignorance score is derived which is unbiased with respect to the ensemble
size.  In an application to seasonal climate predictions it is shown that
the standard Ignorance score assigns better expected scores to simple
climatological ensembles or biased ensembles that have many members, than to
physical dynamical and unbiased ensembles with fewer members.  By contrast,
the new bias-corrected Ignorance score ranks the physical dynamical and unbiased ensembles better than the
climatological and biased ones, independent of ensemble size.  It is shown
that the unbiased estimator has smaller estimator variance and error than the
standard estimator, and that it is a fair verification
score, which is optimized if the ensemble members are
statistically consistent with the observations. The finite
ensemble bias of ensemble verification scores is discussed more broadly. It is
argued that a bias-correction is appropriate when forecast systems with
different ensemble sizes are compared, and when an evaluation of the underlying
distribution of the ensemble is of interest; possible applications to unbiased
parameter estimation are discussed.

\end{abstract}

\section{Introduction}\label{sec:intro}

Weather and climate services routinely issue their forecasts as ensemble
forecasts, i.e.\ collections of forecasts that refer to the same target, but
that differ in their initial conditions, boundary conditions, or model
formulation \citep{sivillo1997ensemble}. Ensembles can serve as the basis to
derive different forecast products, such as point forecasts, using e.g. the
ensemble mean, or probability forecasts, using e.g. the ensemble mean and
standard deviation to forecast a Normal distribution \citep{zhu2005ensemble}.
These different forecast products derived from ensembles
require different methods of forecast verification \citep[ch.
8]{jolliffe2012forecast}. In this paper we shall be particularly interested in the application
of probabilistic scoring rules to ensemble forecasts
\citep{gneiting2007strictly, winkler1996scoring}.

The Ignorance score \citep{roulston2002evaluating}, also called the Logarithmic Score \citep{good1952rational, gneiting2007strictly}, is a proper verification score for probability forecasts. If the forecast is issued as a (unit-less) probability density function $p(z)$ and the forecast target materializes as the value $x$, then the Ignorance score is given by the negative logarithm of the forecast density evaluated at $x$:
\begin{equation}
\mathcal{I}(p; x) = -\log p(x).\label{eq:ign}
\end{equation}
The Ignorance difference between two forecasts $\Delta = -\log q(x) + \log
p(x)$ can be interpreted that the density that $p(z)$ assigns to the
observations $x$ is $e^\Delta$ times as large as the density that $q(z)$
assigns to the same observation $x$. In the negative-log representation of
\refeq{eq:ign}, the Ignorance score acts as a penalty which a forecaster will
try to minimize. When
the natural logarithm is used (as in \refeq{eq:ign}), Ignorance differences are
measured in {\it nats}, and can be transformed to {\it bits} by dividing by
$\log2$, and to {\it bans} by dividing by $\log10$ \citep[sec.
18.3]{mackay2003information}. The Ignorance score has been used as
a verification measures for probabilistic forecasts of weather and climate
\citep{barnston2010verification,krakauer2013information,smith2014probabilistic,
rodrigues2014seasonal}, and for parameter estimation in dynamical systems
\citep{du2012parameter}. The Ignorance score has an information-theoretic
interpretation \citep{roulston2002evaluating, peirolo2011information}, and an
interpretation in terms of betting returns \citep{hagedorn2009communicating}.
\citet{benedetti2010scoring} shows that ``the logarithmic score is the only
[verification score] to respect three basic desiderata whose violation can
hardly be accepted'' and argues that the Ignorance score is therefore the
``univocal measure of forecast goodness''.

If the forecast density is issued as a Normal distribution with mean $\mu$
and variance $\sigma^2$, then the Ignorance is given by
\begin{equation}
\mathcal{I}\left(\mu, \sigma^2; x\right) = \frac12\log 2\pi + \frac12 \log \sigma^2 + \frac12 \left(\frac{x-\mu}{\sigma}\right)^2,
\end{equation}
which follows from the distribution law of the Normal distribution
\citep{gneiting2005calibrated}. The Ignorance score depends on the spread
$\sigma$ of the forecast distribution and on the squared normalized error $[(x-\mu)/\sigma]^2$ of the forecast
mean. If two probability forecasts have the same squared normalized
error, the one with the smaller spread gets assigned the lower Ignorance score.
Likewise, if two forecast distributions have the same spread, the one with the
smaller squared normalized error has the lower score.

Probability forecasts are often generated by running an ensemble of $m$ simulations of a deterministic model to approximate a forecast distribution \citep{gneiting2005weather}. There are different possibilities to transform a finite ensemble into a continuous forecast distribution \citep[e.~g.][]{broecker2008ensemble, deque1994formulation, gneiting2005calibrated}. One simple possibility is to transform the ensemble forecast with members $\{y_1,\cdots,y_m\}$ into a Normal forecast distribution, whose mean and variance are given by the ensemble mean 
\begin{equation}
\hat{\mu} = \frac1m\sum_{i=1}^m y_i \label{eq:mu.hat}
\end{equation}
and the ensemble variance
\begin{equation}
\hat{\sigma}^2 = \frac1{m-1}\sum_{i=1}^m (y_i - \hat{\mu})^2, \label{eq:sigma.hat}
\end{equation}
respectively.

The estimators $\hat\mu$ and $\hat\sigma^2$ are unbiased, that is
$\mathds{E}(\hat\mu)=\mu$ and $\mathds{E}(\hat\sigma^2)=\sigma^2$ for all $m\ge 2$, where
$\mathds{E}(\cdot)$ denotes the expectation with respect to the underlying distribution from which the ensemble members $\{y_1,\cdots,y_m\}$ were drawn. In other words, the sample estimators
$\hat\mu$ and $\hat\sigma^2$ are, on average, equal to the values $\mu$ and $\sigma^2$ of the
underlying distribution; 
$\hat\mu$ and $\hat\sigma^2$ are therefore unbiased with respect to the ensemble size.

Suppose a forecaster chooses to transform an $m$-member ensemble forecast to a
Normal forecast distribution with mean $\hat{\mu}$ and variance
$\hat{\sigma}^2$. If the forecast target materializes as the value $x$, the
Ignorance score of this forecast is 
\begin{equation}
\mathcal{I}(\hat\mu, \hat\sigma^2; x) = \frac12\log2\pi + \frac12\log\hat\sigma^2 + \frac{(x-\hat\mu)^2}{2\hat\sigma^2}.\label{eq:sample.ign}
\end{equation}
Note that, since the ensemble members $\{y_1,\cdots,y_m\}$ are assumed to be
random variables, the sample mean $\hat\mu$, the sample variance $\hat\sigma^2$,
and the Ignorance score given by \refeq{eq:sample.ign}, are random
variables, too.

In section \ref{sec:debiased} we will show that, even though the estimators
$\hat\mu$ and $\hat\sigma^2$ are unbiased with respect to the ensemble size,
the Ignorance score estimated by $\mathcal{I}(\hat\mu, \hat\sigma^2;x)$ is biased, that
is
\begin{equation}
\mathds{E}[\mathcal{I}(\hat\mu, \hat\sigma^2; x)] \neq \mathcal{I}(\mu, \sigma^2; x), \label{eq:unequal}
\end{equation}
where $x$ is assumed constant, and the expectation is taken over the random
variables $\hat\mu$ and $\hat\sigma^2$.  The Ignorance score estimated for a
finite ensemble by \refeq{eq:sample.ign} is, on average, different from the
Ignorance score that the underlying Normal distribution $\mathcal{N}(\mu,
\sigma^2)$ would achieve, if it were known. A finite
ensemble only allows for an imperfect estimation of the underlying
distribution. Therefore, the Ignorance score of the estimated distribution
$\mathcal{N}(\hat\mu, \hat\sigma^2;x)$ is different from the Ignorance score of
the underlying distribution $\mathcal{N}(\mu, \sigma^2; x)$ - different for
particular realisations of a finite ensemble, but also different in
expectation. In this article we will point out a number of consequences of this
inequality, and argue that there are situations where an equality in
\refeq{eq:unequal}, that is, an unbiased estimation of the Ignorance score of
the underlying distribution $\mathcal{N}(\mu, \sigma^2)$, is actually
desirable.

\begin{figure*}
\centering\includegraphics{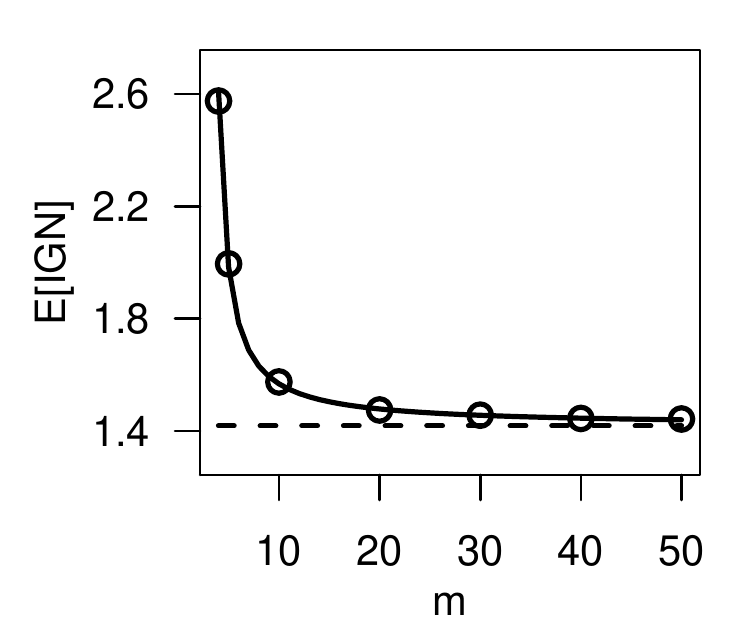}
\caption{Artificial ensembles of size $m$ and observations are drawn independently from a standard Normal distribution. The circles depict the ensemble Ignorance score (calculated by \refeq{eq:sample.ign}), averaged over $10^5$ realizations for each value of $m$. The solid line depicts the mathematical expectation of the ensemble Ignorance score, calculated by the results of section \ref{sec:debiased}. The dashed line depicts the expected Ignorance score of the underlying standard Normal distribution.}\label{fig:m-dependence}
\end{figure*}

\reffig{fig:m-dependence} illustrates the finite-ensemble bias of the
Ignorance score by an artificial example.  Suppose ensembles of size $m$ and
observations are drawn from a standard Normal distribution $\mathcal{N}(0,1)$.  From each
$m$-member ensemble, a Normal forecast distribution $\mathcal{N}(\hat\mu, \hat\sigma^2)$
is constructed, with mean and variance calculated by
\refeq{eq:mu.hat} and \refeq{eq:sigma.hat}.  We have approximated the
expectation of the Ignorance score of the forecasts $\mathcal{N}(\hat\mu,
\hat\sigma^2)$ by simulating $10^5$ ensemble-observation pairs for a few values
of $m$. We have also calculated the expectation analytically, anticipating results from section
\ref{sec:debiased}.  \reffig{fig:m-dependence} shows that the expected
Ignorance score of the ensemble-based forecast $\mathcal{N}(\hat\mu, \hat\sigma^2)$
differs significantly from the expected Ignorance score of the underlying
distribution $\mathcal{N}(0,1)$. The difference is especially large for small ensembles;
for 5-member ensembles, for example, the scores differ by more than $0.5$ in
absolute value, that is, the average score of distributions derived from 5-member ensembles is almost $40\%$ larger than the average score of the underlying distribution. The average Ignorance difference can also be interpreted as an average information deficit of $0.5\ nats$ (or $0.72\ bits$) of the distribution derived from a finite 5-member ensemble compared to the underlying distribution.

The finite ensemble bias of the Ignorance score, its correction, and its
implications for ensemble verification are the main subjects of this paper. The
impact of ensemble-size on forecast performance was studied for example by
\citet{buizza1998impact}, who found that increasing the ensemble size improves
a number of verification measures. The effect of ensemble-size on probabilistic
verification measures, as well as possibilities to quantify or remove the finite ensemble effect, were studied in more detail, for example by
\citet{ferro2007comparing} for the Brier Score, by \citet{ferro2008effect} for
the discrete and continuous ranked probability score, by
\citet{mueller2005debiased} for the ranked probability skill score, and by
\citet{richardson2001measures} for the reliability diagram, the Brier (Skill)
score and potential economic value.  Further discussions of finite-sample
effects on verification scores for ensemble forecasts can be found for example
in \citet{fricker2013three} and \citet{ferro2013fair}.

In section \ref{sec:debiased} of this article an analytic expression of the
finite ensemble bias of the Ignorance score of Normal distributions is derived, as well as a new
estimator of the Ignorance score, which is unbiased with respect to the
ensemble size. The expectation of the new estimator is independent of the
number of ensemble members, and it is an unbiased estimator of the Ignorance score of the underlying distribution of the ensemble. In section \ref{sec:application}
the possible benefits of using a bias-corrected score are illustrated using
data from a seasonal hindcast experiment of average European summer
temperatures. It is shown that the standard Ignorance score favors simple climatological or biased ensemble forecasts with many members over physical dynamical and unbiased ensemble forecasts having fewer members. The new bias-corrected score ranks the physical dynamical and unbiased ensembles better on average, independent of ensemble size.  In section \ref{sec:discussion} we discuss the important difference between the underlying distribution and distributions derived from finite ensembles. We discuss applications where a correction of the finite-ensemble bias of verification scores is desirable. We point out the variance and error reduction of the new Ignorance estimator, consider its applicability to non-Normal ensemble data, and examine the relation to recently proposed fair scores for ensemble forecasts. Section \ref{sec:summary} concludes
with a summary and outlook.

\section{The bias-corrected Ignorance score}\label{sec:debiased}

For the rest of the paper we will refer  to $\mathcal{I}(\mu, \sigma^2; x)$ as the {\it
population Ignorance score} -- it is the score that an infinitely large ensemble drawn from $\mathcal{N}(\mu, \sigma^2)$ would achieve. We will further refer to $\mathcal{I}(\hat\mu, \hat\sigma^2;x)$ as the {\it
standard Ignorance score}, also denoted by $\widehat{\mathcal{I}}$, as it appears to be the natural Ignorance score to calculate for a Normal distribution derived from an ensemble forecast. We remind the reader that simply fitting a Normal distribution to an ensemble forecast might not be the optimal method of deriving a probability distribution from a finite ensemble, and other methods, for example based on kernel dressing \citep{broecker2008ensemble}, might be more applicable. The theory developed in this paper does not apply to such methods. The finite-ensemble bias of $\widehat{\mathcal{I}}$ will be
calculated explicitly in this section, and a {\it bias-corrected Ignorance
score} $\mathcal{I}^*$ for finite ensemble forecasts is derived.

Under the assumption that the ensemble members $\{y_1,\cdots,y_m\}$ are
independent and identically distributed (iid) draws from a Normal distribution
$\mathcal{N}(\mu, \sigma^2)$, the sampling distributions of $\hat{\mu}$ and
$\hat\sigma^2$, as calculated by \refeq{eq:mu.hat} and
\refeq{eq:sigma.hat}, are given by 
\begin{equation}
\hat{\mu} \sim \mathcal{N}\left(\mu, \frac{\sigma^2}{m}\right)\label{eq:mu.hat.dist}
\end{equation}
and
\begin{equation}
\frac{m-1}{\sigma^2} \hat{\sigma}^2 \sim \chi^2_{m-1},\label{eq:sigma.hat.dist}
\end{equation}
where $\chi^2_{m-1}$ denotes the $\chi^2$-distribution with $m-1$ degrees of
freedom; furthermore, $\hat\mu$ and $\hat\sigma^2$ are statistically
independent \citep[sec. 4.3]{mood1950introduction}.

To calculate (and eventually remove) the bias of $\widehat{\mathcal{I}}$, we calculate the
expected values of $\log\hat\sigma^2$ and $(\hat\mu-x)^2/\hat\sigma^2$ under the above assumptions. In
appendices \ref{sec:e.log.sigma} and \ref{sec:e.z2} it is shown that these expectations are 
\begin{align}
\mathds{E} \left[\log \hat\sigma^2\right] = \log\sigma^2 + \Psi\left(\frac{m-1}{2}\right) - \log\left(\frac{m-1}{2}\right),\label{eq:log.sigma.hat}
\end{align}
and
\begin{align}
\mathds{E}\left[\frac{(\hat{\mu}-x)^2}{\hat\sigma^2}\right] = \frac{m-1}{m-3}\left(\frac{\mu-x}{\sigma}\right)^2 + \frac{m-1}{m(m-3)},\label{eq:z.hat.square}
\end{align}
where $\Psi(x)$ is the digamma function\footnote{Numerical approximations of the digamma function are widely implemented in scientific software, for example {\tt digamma(x)} in R (version 3.1.1), and {\tt special.psi(x)} in SciPy (version 0.14.0).}. Note that \refeq{eq:z.hat.square} only holds for $m\ge 4$; otherwise the expectation is undefined due to the diverging second-moment of the t-distribution (cf. appendix \ref{sec:e.z2}).

It follows from \refeq{eq:log.sigma.hat} and \refeq{eq:z.hat.square} that the bias of the standard Ignorance score is given by
\begin{align}
\mathds{E}\left[\mathcal{I}(\hat\mu, \hat\sigma^2; x) - \mathcal{I}(\mu, \sigma^2; x)\right]  = & \frac12\left\{\Psi\left(\frac{m-1}{2}\right) - \log\left(\frac{m-1}{2}\right)\right\}\nonumber\\ 
&  + \frac{1}{m-3}\frac{(x-\mu)^2}{\sigma^2} + \frac{m-1}{2m(m-3)}.\label{eq:bias.ign.hat}
\end{align}
The expectation $\mathds{E}[\cdot]$ is taken over $\hat\mu$ and $\hat\sigma^2$; the
observation $x$ is a constant.  \refEq{eq:bias.ign.hat} shows that, for
finite $m$, the expected standard Ignorance score is different from the
population Ignorance score. 

By combining \refeq{eq:log.sigma.hat} and \refeq{eq:z.hat.square}, and
solving for the population Ignorance score, we find that the score
\begin{align}
\mathcal{I}^*(\hat\mu, \hat\sigma^2; x) = & \frac12 \log2\pi + \frac12\log\hat\sigma^2 + \frac12 \left(\frac{m-3}{m-1}\right)\frac{(\hat\mu - x)^2}{\hat\sigma^2}\nonumber\\
& -\frac12\left\{\Psi\left(\frac{m-1}{2}\right) - \log\left(\frac{m-1}{2}\right) + \frac1m\right\}\label{eq:ign.star}
\end{align}
is an unbiased estimator of the population Ignorance score, that is
\begin{equation}
\mathds{E}[\mathcal{I}^*(\hat\mu, \hat\sigma^2; x)] = \mathcal{I}(\mu, \sigma^2; x).\label{eq:unbiased}
\end{equation}
We will refer to $\mathcal{I}^*(\hat\mu, \hat\sigma^2;x)$ as the {\it bias-corrected
Ignorance score}.  Note that, $\Psi(x) - \log(x)$ is of order $1/x$ for large
$x$ \citep[eq. 6.3.18]{abramowitz1972handbook}. Consequently, $\mathcal{I}^*(\hat\mu,
\hat\sigma^2;x)$ converges to $\mathcal{I}(\mu,\sigma^2;x)$ for $m\rightarrow\infty$.
Moreover, note that unbiasedness implies that the Ignorance score calculated
for a finite ensemble using \refeq{eq:ign.star} is, on average, equal to
the Ignorance score achieved by an infinitely large ensemble for which $\hat\mu=\mu$ and $\hat\sigma^2=\sigma^2$.

\begin{figure*}
\centering\includegraphics{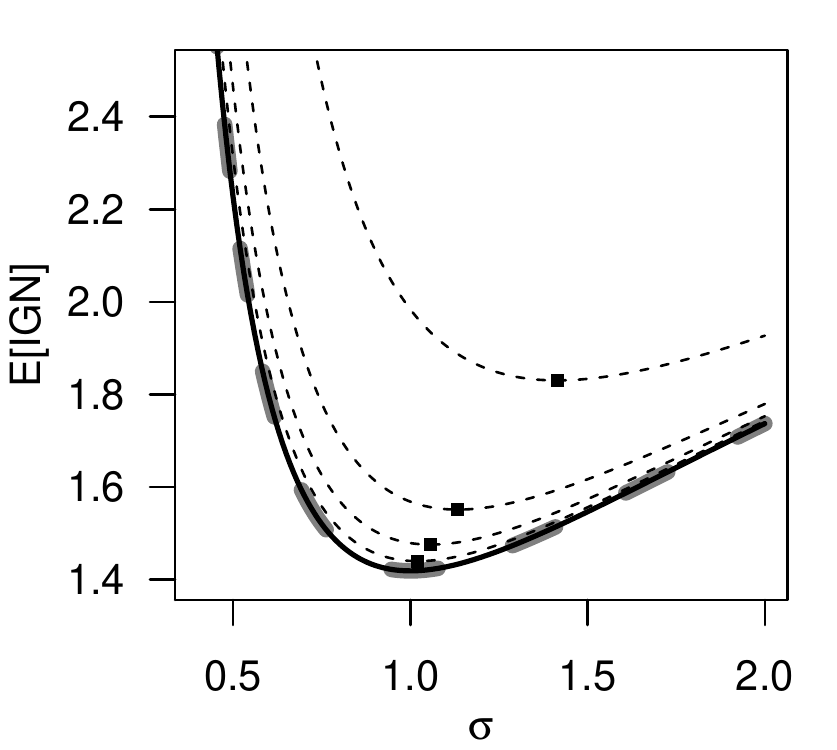}
\caption{Verifications $x$ are drawn iid from $\mathcal{N}(0,1)$, and $m$-member ensembles are drawn iid from $\mathcal{N}(0, \sigma^2)$. The dashed gray line corresponds to the expected Ignorance score for $m\rightarrow\infty$, the black dashed lines correspond to the expected standard Ignorance scores of finite ensembles (from top to bottom: $m=5,10,20,50$), and the black full line shows the expected bias-corrected Ignorance score (independent of $m$). Markers indicate the minima along their respective curves.}\label{fig:1}
\end{figure*}

\reffig{fig:1} illustrates differences between the standard and
bias-corrected Ignorance score if ensembles and observations are drawn from
different distributions (unlike \reffig{fig:m-dependence}). Artificial
observations are drawn again iid from $\mathcal{N}(0,1)$, and artificial $m$-member
ensembles are drawn iid from $\mathcal{N}(0, \sigma^2)$. The expectations of the standard Ignorance score, of the population Ignorance score, and of the
bias-corrected Ignorance score, taken over the distributions of the observations and ensembles, are shown as functions of $\sigma$. Note that these expectations could also be approximated by sample averages over large data sets of forecasts and observations drawn from the respective distributions. The systematic
bias due to the finiteness of the ensemble shows as a vertical offset of the
curves.  The vertical offset is the larger, the smaller the ensemble is,
and at any given value of $\sigma$, the expected standard Ignorance score can be improved
by generating a larger ensemble. In contrast to the standard Ignorance score, the
expectation (or long-term average) of the bias-corrected Ignorance score is equal to the
expected population Ignorance score for all values of $\sigma$ and $m$.

\reffig{fig:1} further shows that the standard Ignorance score rewards
ensembles that violate statistical consistency \citep{anderson1996method} (i.e., the statistical indistinguishability between ensemble members and observations). The expected standard Ignorance score obtains its optimum at a value of $\sigma$ which differs from the standard deviation of the observation. The standard Ignorance score therefore rewards ensemble forecasts that have different statistical properties than the observation. The ensemble that optimizes the expected standard Ignorance score is overdispersive, that is its spread is on average higher than that of the observation. The ensemble forecasts that minimise the expected standard Ignorance score would therefore not pass the test for statistical consistency proposed by \citet{anderson1996method}, and the individual ensemble members cannot be interpreted as equally likely scenarios for the observation.

The expectation of the bias-corrected Ignorance score $\mathcal{I}^*$ is equal to the population Ignorance score. The equality holds for all ensemble sizes greater than 3.  Increasing the ensemble size, say from 5 to 10, does not improve the expected value of $\mathcal{I}^*$. As a consequence of unbiasedness with respect to the ensemble size, the score $\mathcal{I}^*$ does not suffer from the bias of the optimum. The expectation of $\mathcal{I}^*$ is
optimized if the ensemble members are drawn from the same distribution as the observation. Ensembles are rewarded for being statistically consistent with the observation. 

The expectation of the estimator $\mathcal{I}^*$ is insensitive to the number of ensemble members. It can therefore be used to compare ensembles of different sizes. But $\mathcal{I}^*$ also estimates the potential Ignorance score of an infinitely large ensemble.  There might be cases where a $m$-member ensemble is available, but the forecaster is interested in the potential score if the ensemble had $M \neq m$ members. For example, he might be interested in the number of members he would have to generate in order to achieve a certain Ignorance score, or whether his $m$-member ensemble forecasting system would outperform a competing $M$-member ensemble if it had the same number of members. In appendix \ref{sec:ign.mM}, we have derived an estimator of the Ignorance score, denoted $\mathcal{I}^*_{m\rightarrow M}$ (\refeq{eq:ign.mM}), which extrapolates the Ignorance score of an $m$-member ensemble to the score that it would achieve if it had $M\neq m$ members. The score $\mathcal{I}^*_{m\rightarrow M}$ is included for completeness, and will not be discussed further in this article.

\section{Application to seasonal climate prediction}\label{sec:application}

\begin{figure*}
\centering\includegraphics{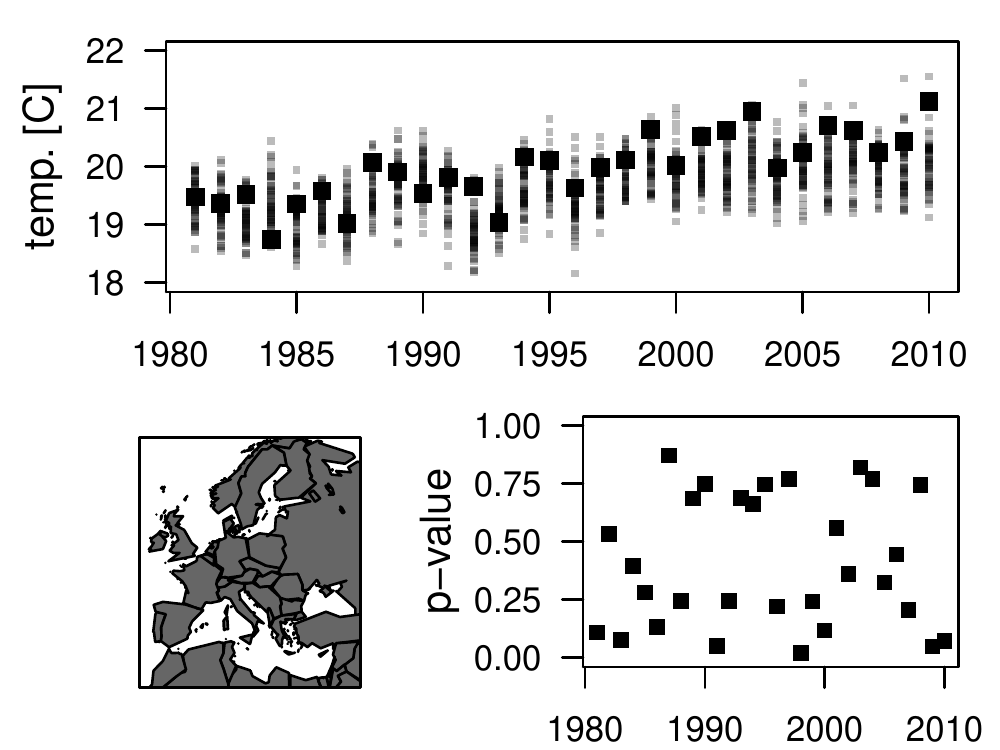}
\caption{Upper panel: Time series of System4 ensemble forecasts (small gray markers) and observations (large black markers) for surface air temperature averaged over the gray area in the lower left panel. Lower right panel: P-values of Shapiro-Wilk normality tests applied to the individual ensemble forecasts, plotted over time.}\label{fig:s4}
\end{figure*}

We illustrate the possible benefits of using a bias-corrected score by a
practical example. We consider ensemble predictions of the summer (JJA) mean
air surface temperature over land over the area limited by 30N -- 75N and 12.5W
-- 42.5E, initialized on the 1 May of the same year. The forecasts are
generated by ECMWF's seasonal forecast system ``System4''
\citep{molteni2011system4} with start dates from 1981 to 2010 ($n=30$), and
$m=51$ ensemble members.  Verifying observations are taken from the WFDEI
gridded data set \citep{weedon2011watch,dee2011era}.  All data were downloaded
through the ECOMS user data gateway \citep{ecoms}. The ensemble and observation
time series are plotted in \reffig{fig:s4}, along with the geographical
region over which temperatures have been averaged.  Visual inspection shows
that a Normal approximation of the ensemble forecasts is justified, which is
strengthened by the approximately uniform distribution of the p-values of
Shapiro-Wilk normality tests applied to the ensembles. The System4 ensemble has a cold bias of $\approx -0.3K$. The observations show a linear trend of $\approx 0.05 K/yr$ which is reasonably well reproduced by the ensemble mean ($\approx 0.03 K/yr$). After removing linear trends, the Pearson correlation coefficient between ensemble means and observations is $0.46$.

We study the effects of finite ensemble sizes by sampling smaller subensembles
from the full 51-member ensemble and calculate their Ignorance scores. At each
time $t=1,\cdots, 30$ we randomly sample $\tilde{m} \le m$ ensemble members
without replacement, and calculate the Ignorance score (averaged over all $t$),
using the estimators $\widehat{\mathcal{I}}$ and $\mathcal{I}^*$. At each value of $\tilde{m}$,
scores are averaged over $10^3$ realizations of random subensembles.

\begin{figure*}
\centering\includegraphics{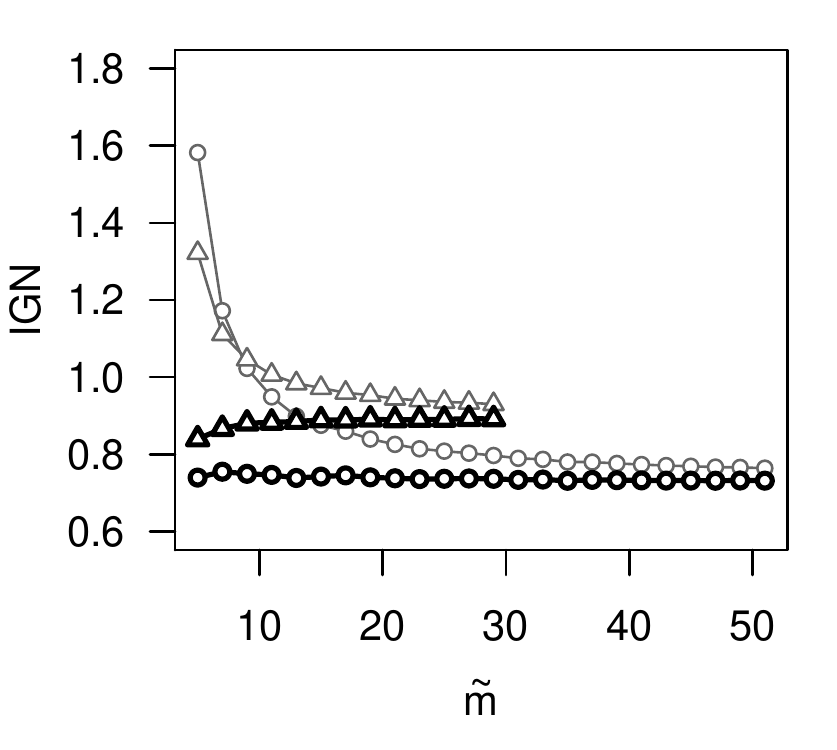}
\caption{Illustration of the effect of finite ensemble sizes on the average Ignorance score, comparing climatological ensembles of size $\tilde{m}$ (triangles) to System4 ensembles of size $\tilde{m}$ (circles). The standard Ignorance score $\widehat{\mathcal{I}}$ is shown in gray, and the bias-corrected Ignorance score $\mathcal{I}^*$ is shown in black.}\label{fig:system4-clim}
\end{figure*}

In \reffig{fig:system4-clim}, the average standard Ignorance score and
average bias-corrected Ignorance score of the $\tilde{m}$-member System4
ensembles are compared to the scores of $\tilde{m}$-member climatological
ensembles. These are randomly sampled without replacement from the 30 years of
observation data.  In order to avoid spurious skill, a climatological ensemble
for time $t$ never includes the observation at time $t$; the maximum value of
$\tilde{m}$ for the climatological ensemble is therefore 29. 
\reffig{fig:system4-clim} shows that the average standard Ignorance score
$\widehat{\mathcal{I}}$ depends systematically on the number of ensemble members, while the
average bias-corrected Ignorance score $\mathcal{I}^*$ is insensitive to the ensemble
size, except for a slight trend at small values of $\tilde{m}$. The dependence
of $\widehat{\mathcal{I}}$ on the ensemble size leads to the conclusion that a 29-member
climatological ensemble is preferable to a 10 member System4 ensemble.  For very small ensemble sizes, the
climatological ensemble has a lower standard Ignorance score than the System4
ensemble, even if the number of members is equal. This difference might be due to the cold bias of the System4 ensemble.
The above remarks highlight the important difference between the two scores: While the standard Ignorance score evaluates the forecast that was derived from a finite ensemble, the bias-corrected Ignorance score evaluates the underlying distribution from which the finite ensemble was drawn. The sensitivity to ensemble size of the score that evaluates the derived forecast can lead to the conclusion that forecasts derived from a large ensemble generated by a simple forecasting system such as climatology is superior to a small ensemble generated by a sophisticated physical-dynamical forecasting system. Larger ensembles allow for more robust estimation of the forecast distribution, which is reflected by the finite-ensemble bias of the standard Ignorance score. On the other hand, the underlying distribution is independent of the ensemble size. If the bias-corrected Ignorance score is used to evaluate this underlying distribution, the more sophisticated
System4 always outperforms the climatological ensemble in this score, independent of the number of ensemble
members. This result suggests that the underlying (time-varying) distributions from which System4 samples its ensembles assign, on average, higher probability to the observations than the (time-constant) climatological distribution.

\begin{figure*}
\centering\includegraphics{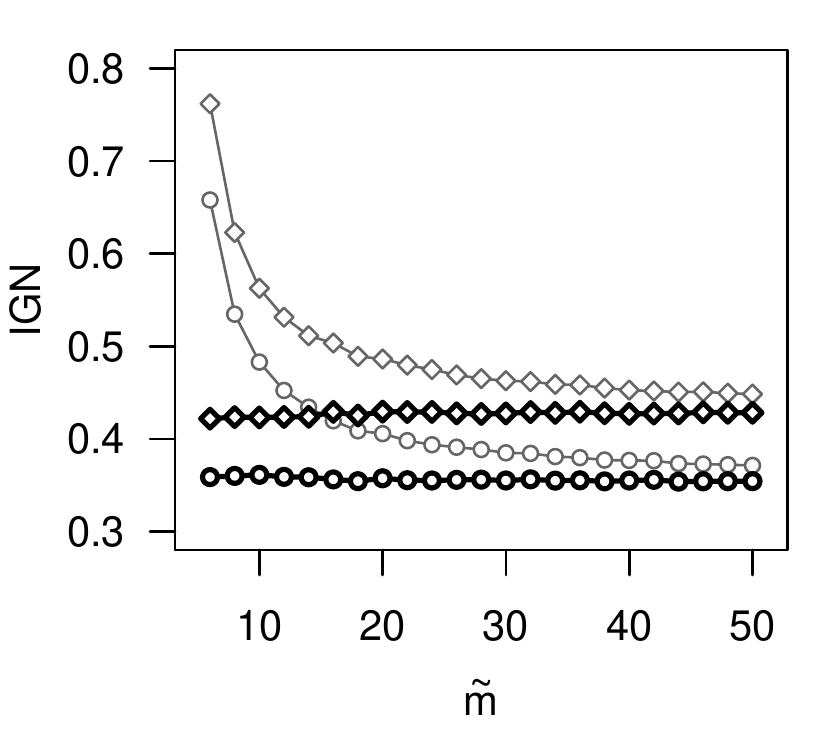}
\caption{Illustration of the effect of finite ensemble sizes on the average Ignorance score, comparing unbiased ensembles of size $\tilde{m}$ (circles) to artificially biased ensembles of size $\tilde{m}$ (diamonds). The standard Ignorance score $\widehat{\mathcal{I}}$ is shown in gray, and the bias-corrected Ignorance score $\mathcal{I}^*$ is shown in black.}\label{fig:system4-bias}
\end{figure*}

For the next analysis we create artificial ensembles whose underlying distributions vary in time, but which are (with great certainty) not more skilful than System4. We first transform System4 ensemble and observation data to anomalies by removing their respective grand averages.  We then create a biased version of System4 by adding to each forecast an artificial bias of 0.25 climatological standard deviations ($\approx 0.15K$). We assume that a verification measure that evaluates the underlying distribution should be expected to rank the unbiased System4 ensemble better than the artificially biased System4 ensemble, independent of ensemble size. In \reffig{fig:system4-bias} we compare the unbiased System4 ensemble with the artificially biased System4 ensemble at different values of $\tilde{m}$, using the standard estimator and bias-corrected estimator of the Ignorance score. \reffig{fig:system4-bias} shows that the standard Ignorance $\widehat{\mathcal{I}}$ on average assigns a better score to a biased ensemble with more than $20$ members than to an unbiased ensemble with less than $10$ members. On the other hand, the bias-corrected Ignorance score always ranks the two ensemble forecasts such that the unbiased ensembles obtains a better average score than the biased ensembles, independent of the number of members. This analysis shows that, forecast distributions derived from a biased ensemble can achieve better average scores than forecast distributions derived from unbiased ensembles if the biased ensemble has more members. The more robust estimation of the forecast distribution from a larger ensemble offsets the disadvantage due to the systematic bias. On the other hand, if the bias-corrected Ignorance score is used to evaluate the underlying distributions of the ensembles, the average score of an inferior (e.g. biased) ensemble cannot be improved simply by adding more members.

The resampling approach of this section can be used as an alternative method to estimating the finite-ensemble effect of a verification score. By subsampling the larger ensemble down to the size of the smaller ensemble, and averaging over many realizations of the subsampling, we can approximate the standard Ignorance score that the larger ensemble would achieve if it had less members. This makes the average standard Ignorance scores of two competing ensembles comparable at the smaller ensemble size. This alternative approach to accounting for the finite-ensemble bias deserves further attention, but will not be studied here in more detail. We only note that subsampling and averaging is computationally more expensive than calculating a bias-corrected score that accounts for the finite-ensemble effect by its mathematical properties. Furthermore, the subsampling approach cannot be used to estimate the score that an ensemble would achieve if it had more members.

\section{Discussion}\label{sec:discussion}

\subsection{ The underlying distribution and derived forecasts}

By correcting the finite-ensemble bias of the standard Ignorance score, we have
derived an unbiased estimator of the score of the underlying distribution.
Deriving a score whose expectation is independent of the ensemble size seems
irreproachable, and has been achieved for different scores (cf. references in
section \ref{sec:intro}).  But the concept of evaluating the unknown underlying
distribution warrants further discussion.  In this section, we argue that a
clear distinction must be made between an evaluation of a derived forecast
distribution, which depends on ensemble size, and an evaluation of the
underlying distribution, which should be independent of ensemble size.

The underlying distribution is a hypothetical concept. 
The individual ensemble members are perceived as random quantities and this randomness is described by an underlying distribution. 
Ensemble forecasting systems can be imagined as (high-dimensional) random number generators that draw samples from the underlying distribution. 
The underlying distribution can therefore be thought of as a part of the ensemble forecasting system. Estimating the score of the underlying distribution can therefore be interpreted as a quantification of the skill of the ensemble forecasting system. 
Evaluating the score of the underlying distribution is therefore especially relevant for model developers, who are interested in the quality of the forecasting system.

Even though we can retrospectively estimate its score, the underlying distribution is not accessible for forecasting.
Only finite ensembles drawn from that distribution are ever available and the number of members is limited by the computational resources. 
Forecasters and decision-makers have to rely on probabilistic forecasts derived from the finite ensemble.
These forecast users will probably be more interested in an estimate of the score of the derived forecast than in an estimate of the score of the underlying distribution.
For these users, the finite-ensemble bias is a practically relevant effect -- more ensemble members allow for a more robust estimation of the forecast distribution, and it is sensible to assume that larger ensembles provide better forecasts.
However, the same user might also be interested in statistically consistent ensemble forecasts, for example if individual members of a global forecast model are used as scenarios to drive a high-resolution regional model. In that case, rewarding ensembles for being statistically consistent by using a bias-corrected score might be preferable.

We have shown that, due to the finite-ensemble bias, the average Ignorance score of the derived forecast is not necessarily indicative of the average Ignorance score of the underlying distribution.
Due to the finite-ensemble bias, it is possible that a forecast derived from a statistically inconsistent, biased, or simple climatological forecast system outperforms forecasts derived from statistically consistent, unbiased, or sophisticated physical-dynamical ensemble forecasts.
If the underlying distribution were available, the latter forecasts might be preferable to the former. But the difference in ensemble size, and the fact that derived forecasts from finite ensembles are evaluated, reverses the preference.

\subsection{When is the finite-ensemble bias-correction desired?}

There are situations, where the bias-correction is clearly not desired, namely
when the quality of the derived probability forecast is of interest, instead of
the quality of the underlying ensemble distribution. If a probability forecast
is always generated using a specified number of ensemble members, and a Normal
approximation $\mathcal{N}(\hat\mu, \hat\sigma^2)$ is used, the standard Ignorance score $\widehat{\mathcal{I}}$ rather than the bias-corrected Ignorance score $\mathcal{I}^*$ is the correct
version of the Ignorance score to evaluate this probability forecast. The potential score for
$m\rightarrow\infty$ is of no interest in this case, because only finitely many
ensemble members are ever available. The bias of the optimum is acceptable if
it implies that a statistically inconsistent ensemble provides a better
probability forecast than a statistically consistent one. However, due to the bias of the optimum, the members of the optimal
ensemble are not necessarily exchangeable with the observation. The individual
members must therefore not be interpreted as ``possible future scenarios''. In
fact, the individual ensemble members should not be interpreted at all in this
case; only the derived continuous forecast distribution is of interest.

There are at least three applications where an estimation or
correction of the finite-ensemble bias is clearly desirable: Firstly,
in numerical model development, often new ensemble
prediction systems are to be explored, using for example a new initialization
technique, a new parametrization scheme, or an experimental dynamical core.
If we adopt the notion that the hypothetical underlying distribution
is a property of the forecasting system, such modifications of the forecasting
system change, and possibly improve, the score of the underlying distribution.
In such pilot studies, it might be desirable to limit CPU time by generating
ensembles with fewer members than the final (operational) forecast product will
have, and then accounting for the finite ensemble bias
by using a suitable score. Estimating the score of the underlying distribution then provides
a more realistic estimate of the score of the final product, especially
in relation to competing forecasting systems with possibly larger ensembles.

Secondly, if ensembles of different sizes are compared and forecasts
derived from them have different scores, it might be of interest whether the
larger ensemble achieves a
better score due to the finite-ensemble bias, or because its underlying distribution is
more skilful at predicting the real world and therefore assigns higher probablity to the observations.  This difference is illustrated in
section \ref{sec:application}, where it is shown that a biased ensemble can
outperform an unbiased ensemble merely by having more members. A score that accounts for the finite-ensemble bias can be used to inform the forecaster that increasing the size of the smaller ensemble to the size of the bigger ensemble could produce an even better forecast.

Lastly, the bias of the optimum of the standard Ignorance
score, illustrated in \reffig{fig:1}, is clearly undesirable when
the Ignorance score is used as an objective function for parameter estimation
and optimization. For example, a verification score might be used to tune parameters of
the ensemble forecasting system to match as well as possible the corresponding
parameter of the observation (such as the standard deviation in
\reffig{fig:1}). A biased score which evaluates the derived forecast distribution might favor an ensemble that differs
systematically from the observations. Even though the ensemble system optimized by the biased score has been tuned to generate the best possible forecasts, the ensemble members do not behave like the real world. Removing the bias of the optimum by using
a bias-corrected score ensures that the optimised ensemble is statistically
consistent with the observations. The optimised parameter values of the ensemble are equal to the parameter values of the real world (provided the parameter really has a physical interpretation). Additionally, the optimal value of the parameter does not change if the ensemble size is changed. If the score that is used for parameter tuning has a bias of the optimum, the parameters would have to be re-tuned whenever the ensemble size is changed. We consider unbiased parameter estimation and optimisation a relevant and promising application of bias-corrected verification scores, but more work is necessary to fully explore their applicability and limitations.

\subsection{ Propriety and fairness}

A scoring rule $S$ is proper (relative to a class $\mathcal{P}$ of probability measures) if, for any $q\in\mathcal{P}$, the expected score taken over the distribution $q$ of the observation $x$, $\mathds{E}_{x\sim q}[S(p,x)]$, is optimized when $p=q$ \citep{gneiting2007strictly}.
A proper scoring rule thus favours (on average) forecasts $p$ that equal the distribution $q$ of the observation.
The standard Ignorance score defined by \refeq{eq:ign} is a proper scoring rule.
Here, $S$ is a function of an observation $x$ and a {\it distribution} $p$.
Our forecast is an ensemble, not a distribution, so we must decide what distribution to use for $p$.
If we use a distribution derived from the ensemble, such as $\mathcal{N}(\hat\mu, \hat\sigma^2)$, then the standard Ignorance score favours (on average) ensembles for which the derived distribution $\mathcal{N}(\hat\mu,\hat\sigma^2)$ equals the distribution of the observation.
If we want to evaluate the derived distribution as a probability forecast then the standard Ignorance score is a proper score to use.
But we have shown that ensembles that optimise the standard Ignorance score on average are not those whose {\it underlying} distribution equals the distribution of the observation.

Recently, \citet{fricker2013three} and \citet{ferro2013fair} introduced fair scores as a possible extension of proper scores to ensemble forecasts. If we want to evaluate the underlying distribution, and thereby favour ensembles whose underlying distribution equals the distribution of the observation (so that ensembles and observations are statistically consistent, for example) then we should use a fair score.  A scoring rule $S^*$ is fair (relative to a class $\mathcal{P}$ of probability measures) if, for any $p,q\in \mathcal{P}$, the expectation of the score, taken with respect to the distribution $q$ of the observation $x$, and with respect to the distribution $p$ from which the ensemble members $y_i$ were independently drawn, $\mathds{E}_{x \sim q}\mathds{E}_{y_i \sim p}[S^*(y,x)]$, is optimized when $p=q$.  A fair score thus favours (on average) ensembles whose underlying distribution is equal to the distribution of the observation.  The bias-corrected Ignorance score is fair relative to the class of Normal distributions. Here, $S^*$ is a function of an observation $x$ and an {\it ensemble} $y$.  Since $y$ is not a distribution, it would not be meaningful to ask whether $\mathcal{I}^*$ is proper.  It is true, however, that $S(p,x) := \mathds{E}_{y_i \sim p}[S^*(y,x)]$ is proper. The reader is referred to \citet{ferro2013fair} for more discussion of the relation between fair and proper scores.

Unbiasedness of the bias-corrected Ignorance $\mathcal{I}^*$ only
holds for independent and identically Normal distributed ensemble
members.  If the ensemble members are non-Normal, $\mathcal{I}^*$ is
biased (cf. section \ref{sec:non-normal} and appendix \ref{sec:violations}).
Therefore, (adopting the terminology of \citet{gneiting2007strictly}),
we say that the score $\mathcal{I}^*$ is fair
relative to the class of Normal distributions.
In contrast, the fair continuous ranked probability score (fair CRPS) for
continuous ensemble forecasts proposed by \citet{fricker2013three} is
independent of the distribution of the ensemble members, and therefore has
wider applicability than the fair Ignorance score presented here. Note that
there is an interesting difference between the Ignorance score and the CRPS.
According to \reffig{fig:1}, the standard Ignorance
score favors overdispersive ensemble forecasts. By contrast, according to
Figure 2 of \citet{ferro2013fair}, the CRPS without fairness adjustment favors
underdispersive ensembles. This difference shows that without a bias-correction
for finite ensemble sizes, different proper scores can favor ensembles that are
not only inconsistent with the observation, but the nature of the inconsistency
can also be fundamentally different for different scores. This is shown by the
bias of the optimum of the standard deviation, which is positive for the
standard Ignorance score and negative for the unadjusted
CRPS.

\subsection{ Non-Normal data}\label{sec:non-normal}

We have shown in section \ref{sec:debiased} that the bias-corrected Ignorance
score $\mathcal{I}^*$ completely removes the finite-ensemble bias if the ensemble
members are identically and independently Normal distributed. In practical
applications such as atmospheric forecasts, where ensemble members are
generated by complex numerical computer simulations, Normality appears to be
too strong an assumption. It is unrealistic to assume that outputs from
computer simulations are exactly Normally distributed. However, if the ensemble
members are not iid Normal distributed, a basic assumption in the derivation of
$\mathcal{I}^*$ is violated, and $\mathcal{I}^*$ might be biased after all.

We show in appendix \ref{sec:violations} that for non-Normal ensemble members,
$\mathcal{I}^*$ is indeed biased. This is shown for ensembles with heavy-tailed
distributions, skewed distributions and bimodal distributions. But the bias of
$\mathcal{I}^*$ is always considerably smaller than the bias of $\widehat{\mathcal{I}}$. This
reduction of the finite-ensemble bias implies that if the ensemble data
suggests a Normal approximation, and if the finite-ensemble bias of the
Ignorance score is undesired, $\mathcal{I}^*$ should be used for ensemble verification,
rather than $\widehat{\mathcal{I}}$.

\subsection{Bias-variance decomposition}

Bias is not the only factor that contributes to differences between a finite
sample estimator and the corresponding population value. Another important
factor is the estimator variance, i.e.\ the average squared difference of the
estimator from its expectation. The sum of the squared bias and the
variance can be shown to be equal to the expected squared error of the
estimator, i.e.\ the expected squared difference between the estimator and the
population value \citep[sec. 7.3]{mood1950introduction}. That is, an unbiased
estimator can still have a larger error than a biased estimator, by having a
very large variance. This is not the case for $\mathcal{I}^*$. We show in appendix \ref{sec:var.ign},
that under a first order approximation, the conditional variance of
$\mathcal{I}^*(\hat\mu, \hat\sigma^2;x)$, given $x$, is always smaller than the
conditional variance of $\widehat{\mathcal{I}}$. This means, that $\mathcal{I}^*$ is not only equal
to the population score on average, it is also on average closer to the
population score in a mean-squared sense, which provides further motivation to
use $\mathcal{I}^*$ instead of $\widehat{\mathcal{I}}$.

\section{Summary and outlook}\label{sec:summary}

We have studied the applicability of the Ignorance score for ensemble
verification. We focused on Normal approximations, where the ensemble forecast
is transformed to a continuous Normal forecast distribution whose parameters
are estimated by the ensemble mean and variance. It was shown that the
Ignorance score applied to this forecast distribution is biased with respect to
the ensemble size: Larger ensembles obtain systematically better scores. In
section \ref{sec:debiased} a new estimator of the Ignorance score was derived
which removes the finite ensemble bias; the expectation of the new
estimator is independent of the number of ensemble members. The main advantage
of the new score is that it allows for a fair comparison of ensemble forecasts
with different number of members. This was illustrated in section
\ref{sec:application} by application to seasonal climate forecasts.  It was
shown that the standard Ignorance score favors
biased or climatological ensembles with many members
over unbiased and physical dynamical
ensembles with few members. In contrast, the bias-corrected Ignorance score
on average ranks the statistically
consistent and physical dynamical ensemble better, regardless of the number of members.  In section
\ref{sec:discussion}, we concluded that the bias-corrected Ignorance score is
applicable also to non-Normal ensemble data, and that the new score estimator
not only reduces the bias, but also the estimator variance, thereby decreasing
the overall estimation error of the score. It was shown that the new estimator
is a strictly fair score, and situations were discussed when bias-corrected scores are preferable for ensemble evaluation.  

There is some scope to extend the results of this paper. For example, it might
be possible to derive a bias-corrected Ignorance score of a multivariate Normal
forecast, since the sampling distributions of the multivariate sample mean and
covariance matrix are known. Further, a bias-corrected score for affine
transformations of ensemble forecasts would be useful to assess the quality of
ensemble forecasts that were recalibrated by such a transformation. However,
affine transformations can introduce correlations between the ensemble members, and deriving a bias-corrected score for non-independent ensembles is difficult. It
might also be possible to derive bias-corrected estimators of different
verification scores under a Normal assumption, or bias-corrected estimators of
the Ignorance under different distributions.  This paper provides a framework
for how these problems can be approached, and how the properties of the
resulting estimators can be analysed.

The results of this article are potentially useful outside the area of forecast
verification. First of all, the Ignorance score can be regarded as the negative
log-likelihood of a Normal distribution which is represented by a finite
sample. The bias correction derived in section \ref{sec:debiased} might be
useful for maximum likelihood parameter estimation.  Furthermore, the Ignorance
score is motivated by the entropy as an information-theoretic measure of
predictability \citep{roulston2002evaluating}.  The bias-corrected version
derived in this article might therefore be useful to account for
finite-ensemble effects in information-theoretic predictability frameworks
\citep{delsole2007predictability}.

\section*{Acknowledgments}

We are grateful for stimulating discussions with the members of the statistics
group of the Exeter Climate Systems, in particular Robin Williams, Phil Sansom,
and Keith Mitchell. Comments from two anonymous reviewers helped to considerably improve the paper. This work was funded by the European Union Programme
FP7/2007-13 under grant agreement 3038378 (SPECS).

\begin{appendix}
\section{Appendix: Proofs}

\subsection{$\mathds{E}[\log\hat{\sigma}]$}\label{sec:e.log.sigma}

The derivation follows from the properties of distributions in the exponential family and their sufficient statistics \citep[sec. 1.5]{lehmann1998theory}.  If $X\sim \chi^2_{m-1}$, we can define $\tau:= (m-1)/2 -1$ and write the pdf of $X$ as
\begin{equation}
p_X(x) = \exp\left\{ \tau\log x - \frac{x}{2} - (\tau+1)\log 2 - \log\Gamma(\tau+1) \right\}.\label{eq:chi2.exp}
\end{equation}
Differentiating the integral $\int dx\ p_X(x)$ with respect to $\tau$ yields
\begin{equation}
\mathds{E}[\log X] = \log 2 + \Psi\left(\frac{m-1}{2}\right)\label{eq:e.log.x}
\end{equation}
where $\Psi(x)=d/dx \log\Gamma(x)$ is the digamma function. Applying \refeq{eq:e.log.x} to $\hat\sigma^2$, whose distribution is given by \refeq{eq:sigma.hat.dist}, we get
\begin{align}
\mathds{E} \left[\log \hat\sigma^2\right] = & \mathds{E}\left[ \log \frac{m-1}{\sigma^2} \hat{\sigma}^2\right] + \log \frac{\sigma^2}{m-1}\\
= &  \log\sigma^2 + \Psi\left(\frac{m-1}{2}\right) - \log\left(\frac{m-1}{2}\right).
\end{align}

\subsection{$\mathds{E}\left[\left(\frac{\hat{\mu}-x}{\hat{\sigma}}\right)^2\right]$}\label{sec:e.z2}

Let the independent random variables $Z$ and $V$ have distributions $Z\sim \mathcal{N}(0,1)$ and $V\sim \chi^2_{m-1}$. Then the non-central t-distribution $t_{m-1,x}$, with $m-1$ degrees of freedom and noncentrality parameter $x$, is defined through 
\begin{equation}
\frac{Z+x}{\sqrt{V/(m-1)}} \sim t_{m-1, x}\label{eq:t-def}
\end{equation}
Using the sampling distributions of $\hat\mu$ and $\hat\sigma^2$, and their independence, we get the following relation:
\begin{align}
\sqrt{m}\frac{\hat\mu-x}{\hat\sigma} = & \frac{\frac{\hat\mu-\mu}{\sigma/\sqrt{m}} + \frac{\sqrt{m}}{\sigma}(\mu-x)}{\sqrt{\frac{m-1}{\sigma^2}\hat\sigma^2}/\sqrt{m-1}}\\
\sim\ & t_{m-1, \frac{\sqrt{m}}{\sigma}(\mu-x)}.
\end{align}
The raw moments of a random variable $T\sim t_{m,x}$ are given by \citet{hogben1961moments}:
\begin{equation}
\mathds{E}\left[T^k\right] = \left(\frac{m}{2}\right)^{\frac{k}{2}} \frac{\Gamma\left(\frac{m-k}{2}\right)}{\Gamma\left(\frac{m}{2}\right)} \exp\left( -\frac{x^2}{2}\right) \frac{\partial^k}{\partial x^k} \exp\left(\frac{x^2}{2}\right).
\end{equation}
By calculating the second raw moment of $\sqrt{m}(\hat\mu-x)/\hat\sigma$ and dividing by $m$ we get
\begin{align}
\mathds{E}\left[\left(\frac{\hat{\mu}-x}{\hat{\sigma}}\right)^2\right] = \frac{m-1}{m-3}\left(\frac{\mu-x}{\sigma}\right)^2 + \frac{m-1}{m(m-3)}.
\end{align}

\subsection{$var(\mathcal{I}^*) \le var(\widehat{\mathcal{I}})$}\label{sec:var.ign}

In this appendix, we calculate approximate expressions for the variances of the
standard and bias-corrected Ignorance score. To be more precise, we calculate the
conditional variance of $\mathcal{I}(\hat\mu, \hat\sigma^2; x)$, given $x$. Only
variability of the random variables $\hat\mu$ and $\hat\sigma^2$ contributes to
the variance, while the observation $x$ is kept constant. The results are used
to show that $var(\mathcal{I}^*) \le var(\widehat{\mathcal{I}})$.

It follows from the sampling distributions of $\hat\mu$ and $\hat\sigma^2$
(\refeq{eq:mu.hat.dist} and \refeq{eq:sigma.hat.dist}) that
$var(\hat\mu) = \sigma^2/m$ and $var(\hat\sigma^2) = 2\sigma^4/(m-1)$.
Furthermore, $\hat\mu$ and $\hat\sigma^2$ are statistically independent. We
use these results to approximate the variance of $\widehat{\mathcal{I}}$ by propagation
of error \citep[sec. 2.3]{mood1950introduction}. A first-order Taylor
expansion of $\widehat{\mathcal{I}}(\hat\mu, \hat\sigma^2;x)$ around $\mu$ and
$\sigma^2$ yields
\begin{equation}
var [\widehat{\mathcal{I}}(\hat\mu, \hat\sigma^2;x)] \approx \left(\frac{\partial\widehat{\mathcal{I}}}{\partial\hat\mu} \right)^2 var(\hat\mu) + \left(\frac{\partial\widehat{\mathcal{I}}}{\partial\hat\sigma^2}\right)^2 var(\hat\sigma^2)
\end{equation}
and analogously for $\mathcal{I}^*$. Define the variable $\hat{z} = (\hat\mu - x) / \hat\sigma$. Then the approximate variances of $\widehat{\mathcal{I}}$ and $\mathcal{I}^*$ are given by
\begin{align}
var(\widehat{\mathcal{I}}) & \approx \frac{\hat{z}^2}{m} + \frac{1}{2(m-1)}\left(1 - \hat{z}^2\right)^2\label{eq:var.ign.hat}\text{, and}\\
var(\mathcal{I}^*) & \approx \left(\frac{m-3}{m-1}\right)^2 \frac{\hat{z}^2}{m} + \frac{1}{2(m-1)} \left(1-\frac{m-3}{m-1}\hat{z}^2\right)^2\label{eq:var.ign.star}
\end{align}
It follows from \refeq{eq:var.ign.hat} and \refeq{eq:var.ign.star} that the difference between the variances is
\begin{equation}
var(\widehat{\mathcal{I}}) - var(\mathcal{I}^*) = \frac{2\hat{z}^4}{(m-1)^2} \left(1 - \frac{1}{m-1}\right) +  \frac{2\hat{z}^2}{m(m-1)^2} (m-4)
\end{equation}
which is non-negative for all $m\ge 4$. That is, under the first-order
approximation, the conditional variance of $\mathcal{I}^*$, given $x$, is never
greater than the conditional variance of $\widehat{\mathcal{I}}$. The result is
independent of the value of the observation $x$.

\subsection{$\mathcal{I}^*_{m\rightarrow M}$}\label{sec:ign.mM}

We write the ensemble mean and variance calculated from an $m$-member
ensemble by $\hat\mu_m$ and $\hat\sigma^2_m$, respectively. In this appendix
we show how to use $\hat\mu_m$ and $\hat\sigma^2_m$ to estimate the
Ignorance score that the same ensemble would achieve if it had $M\neq m$
members. First note that it follows from \refeq{eq:log.sigma.hat} that
\begin{align}
\mathds{E}\left[\log\hat\sigma^2_M\right]\nonumber = \mathds{E} \Bigg[ & \log\hat\sigma^2_m - \Psi\left(\frac{m-1}{2}\right) + \Psi\left(\frac{M-1}{2}\right) + \nonumber\\
& \log\left(\frac{m-1}{2}\right) - \log\left(\frac{M-1}{2}\right)\Bigg],\label{eq:mM1}
\end{align}
where $\hat\sigma^2_m$ and $\hat\sigma^2_M$ are ensemble variances of $m$- and $M$-member ensembles sampled from the same distribution. Similarly, it follows from \refeq{eq:z.hat.square} that
\begin{align}
\mathds{E}\left[\frac{(\hat\mu_M-x)^2}{\hat\sigma^2_M}\right] = \mathds{E} \Bigg[ & \left\{ \left(\frac{m-3}{m-1}\right) \frac{(\hat\mu_m-x)^2}{\hat\sigma^2_m} - \frac{1}{m}\right\} \frac{M-1}{M-3} + \nonumber\\
& \frac{M-1}{M(M-3)} \Bigg],\label{eq:mM2}
\end{align}
where $\hat\mu_m$ and $\hat\mu_M$ are ensemble means of $m$- and $M$-member ensembles sampled from the same distribution. Using \refeq{eq:mM1} and \refeq{eq:mM2}, we can derive the score
\begin{align}
&\mathcal{I}^*_{m\rightarrow M}(\hat\mu_m, \hat\sigma^2_m; x)\\ 
& = \frac12\log 2\pi + \frac12 \log\hat\sigma^2_m + \frac12\left(\frac{M-1}{M-3}\right)\left(\frac{m-3}{m-1}\right)\frac{(\hat\mu_m - x)^2}{\hat\sigma^2_m}\nonumber\\
& + \frac12\left[ \Psi\left(\frac{M-1}{2}\right) - \Psi\left(\frac{m-1}{2}\right) \right.\nonumber\\ 
& + \log\left(\frac{m-1}{M-1}\right) + \left.\frac{(m-M)(M-1)}{Mm(M-3)}\right],\label{eq:ign.mM}
\end{align}
which satisfies
\begin{equation}
\mathds{E}\Big[\mathcal{I}^*_{m\rightarrow M}(\hat\mu_m, \hat\sigma^2_m;x)\Big] = \mathds{E}\Big[ \widehat{\mathcal{I}}(\hat\mu_M, \hat\sigma^2_M;x)\Big].
\end{equation}
That is, the score $\mathcal{I}^*_{m\rightarrow M}$ is a function of the sample mean
and variance of an $m$-member ensemble, but on average it is equal to the
Ignorance score of a hypothetical $M$-member ensemble sampled from the same
distribution.

\section{Appendix: Behavior for Non-Normal ensemble data}\label{sec:violations}

In this appendix, we consider the effect of non-Normal ensemble data on the
bias of the Ignorance score.  If the ensemble members are not iid Normal
distributed, an additional systematic error arises. By making the Normal
assumption for non-Normal ensembles, possible features of the forecast
distribution such as heavy-tailedness, skewness, or multimodality are
neglected. Suppose the observation is a skewed random variable, and the
ensemble is indeed drawn from the correct skewed distribution. Transforming the
ensemble to a Normal distribution degrades the skill of the ensemble
forecasting system, because skewness is ignored. In this case, the average
Ignorance score of the Normal approximation of the forecast distribution is
worse than the average Ignorance score that the true ensemble distribution
would achieve, if it were known. Clearly, the ensuing bias due to non-Normality
of the ensemble is not removed by $\mathcal{I}^*$.

For ensemble forecasts which are obviously non-Normal, i.e.\ which have members
that are gross outliers, which are heavily skewed or which exhibit strong
multimodality, a Normal assumption would not be made in practice. The
Ignorance score should not be estimated by \refeq{eq:sample.ign} for
these ensembles, and a bias-corrected Ignorance score for Normal ensembles is of no
interest in such cases.

On the other hand, there might be a moderate violation of Normality, which is
not immediately obvious, or which is small enough such that a Normal assumption
seems a good approximation. In this section, we consider three kinds of
moderate deviations from Normality that might occur in practical applications:
Heavy-tailedness, skewness, and bimodality. In order to keep things simple, we
consider only reliable ensembles which are always drawn from the same
distribution as their verifying observation.  Furthermore, all distributions
are scaled and shifted to have zero mean and unit variance. In each case, the
degree of Normality is tuned by a distribution-specific Normality parameter
$\theta$, that has a limiting value for which the respective distribution
converges to the standard Normal distribution. 

\begin{figure*}
\centering\includegraphics{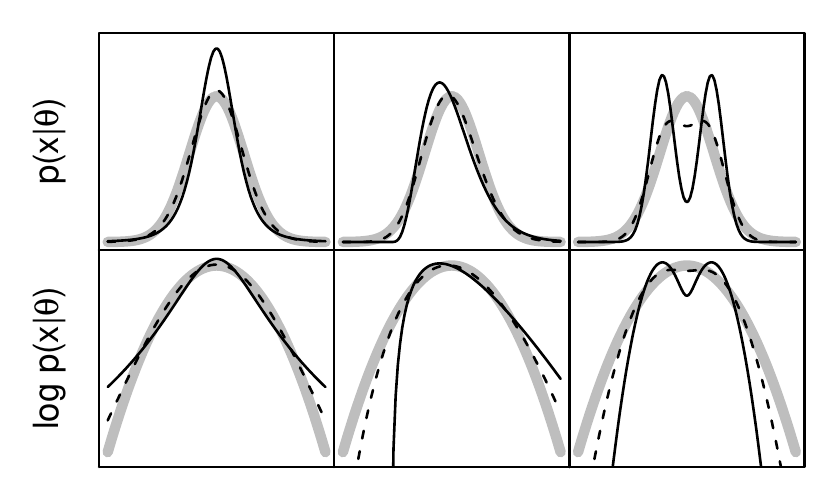}
\caption{Illustration of the types of Non-Normality considered. Left column: Heavy-tailed t-distributions with $\theta=4$ (\solidline) and $\theta=20$ (\dashedline) degrees of freedom. Middle column: Positively skewed Gamma distributions with shape parameter $\theta=5$ (\solidline) and $\theta=50$ (\dashedline). Right column: Bimodal Normal mixture with modes at $\theta=\pm0.9$ (\solidline) and $\theta=\pm0.75$ (\dashedline). All distributions have been scaled and shifted to have zero mean and unit variance. The standard Normal distribution is shown in gray.}\label{fig:nonnormal}
\end{figure*}

We simulate heavy-tailed ensembles and observations by Student's
t-distribution, denoted by $t_\theta$. The parameter $\theta$ (which we assume
to be $>2$) denotes the degree of freedom of the t-distribution. The
t-distribution has zero mean and converges to the Normal distribution for
$\theta\rightarrow\infty$. The random variable $X\sim t_\theta$ has variance
$\theta/(\theta-2)$. Thus, the random variable $\sqrt{(\theta-2)/\theta}X$ has
unit variance, as desired for our study.
Secondly, we simulate skewed ensembles and observations by a Gamma distribution
$\Gamma(\theta, \sqrt{\theta})$ with shape parameter $\theta$. By setting the
rate parameter to $\sqrt{\theta}$ the variance is set to unity. The random
variable $X\sim\Gamma(\theta,\sqrt{\theta})$ has mean equal to $\sqrt{\theta}$,
thus the random variable $X-\sqrt{\theta}$ has zero mean.  The Gamma
distribution $\Gamma(\theta,\sqrt{\theta})$ has skewness $2/\sqrt{\theta}$ and
converges to the Normal distribution for $\theta\rightarrow\infty$.
Lastly, we simulate bimodal ensembles and observations by a Normal mixture.
Define the Bernoulli-distributed random variable $U\sim Ber(1/2)$ and the
Normal random variable $Y\sim \mathcal{N}(\theta, 1-\theta^2)$. Then the random variable
$X=(2U-1)Y$ has a bimodal Normal distribution with modes at $\pm \theta$, zero
mean, and unit variance\footnote{Note that the distribution function is truly
bimodal (i.e.\  has a local minimum at $x=0$) only for $\theta>\sqrt{1/2}$}. The
parameter $\theta\in[0,1)$ tunes the Normality; for $\theta=0$ the distribution
of $X$ is the standard Normal.
The three types of non-Normal distributions are sketched in 
\reffig{fig:nonnormal} for different values of $\theta$.

For observations and ensembles sampled from a Nonnormal distribution with some parameter $\theta$, we calculate 4 different Ignorance scores: 
\begin{enumerate}
\item The population Ignorance score of the underlying distribution
$p(x|\theta)$ from which the observation was sampled, denoted by $\mathcal{I}_p:=-\log
p(x|\theta)$.
\item The Ignorance score of the Normal approximation of $p(x|\theta)$, denoted
by $\mathcal{I}_n$. Recall that all non-Normal distributions that we consider always
have zero mean and unit variance; the standard Normal is therefore always the
best Normal approximation of the true $p(x|\theta)$, and we have
$\mathcal{I}_n(x)=\log(2\pi)^{-1/2}+x^2/2$ every time. 
\item The standard Ignorance score $\widehat{\mathcal{I}}$, calculated by 
\refeq{eq:sample.ign}, using the ensemble mean and ensemble standard deviation;
and 
\item The bias-corrected Ignorance score $\mathcal{I}^*$, calculated by 
\refeq{eq:ign.star}, where we also use the ensemble mean and ensemble standard
deviation. 
\end{enumerate}
Note that $\mathcal{I}_p$ and $\mathcal{I}_n$ are independent of any ensemble forecast.

\begin{figure*}
\centering\includegraphics{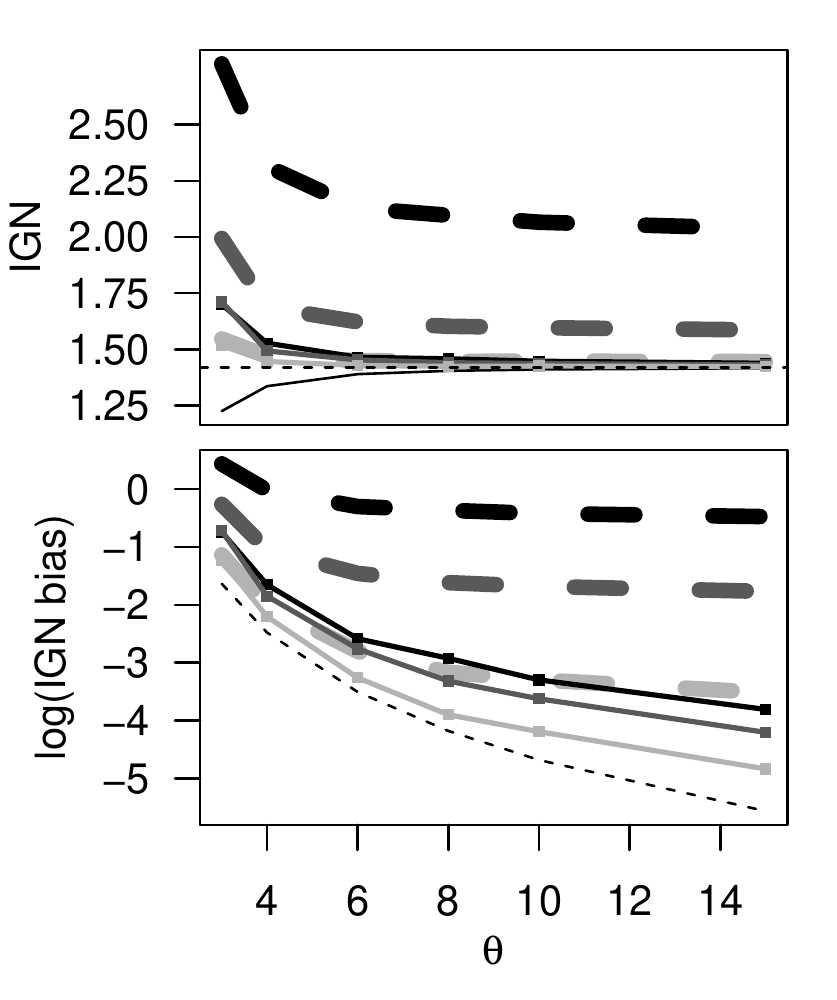}
\caption{Ignorance scores and their biases for heavy-tailed data as functions
of the non-Normality parameter $\theta$. Upper panel: Population Ignorance
scores of the t-distribution (\solidline), of the Normal approximation $\mathcal{N}(0,1)$
(\dashedline), and for finite ensemble forecasts (black: $m=5$, dark gray:
$m=10$, light gray: $m=50$; thick dashed lines: standard Ignorance score, thinner
lines with markers: bias-corrected Ignorance score). Lower panel: The
respective biases, i.e.\ the differences between the three Ignorance scores that
use a Normal approximation, and the population Ignorance score of the
t-distribution, plotted in log-normal axes. See text for details about the
scores and the definition of the non-Normality parameter
$\theta$.}\label{fig:nonnormal-bias-t}
\end{figure*}

\begin{figure*}
\centering\includegraphics{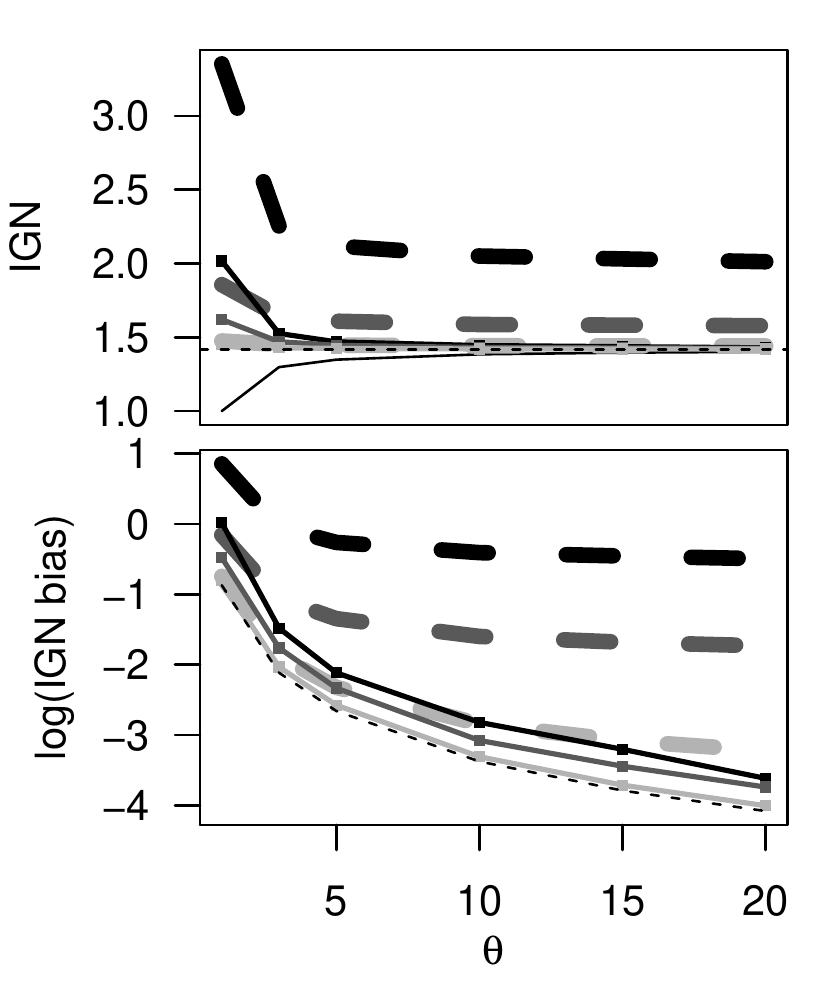}
\caption{Same as \reffig{fig:nonnormal-bias-t} but for skewed distributions. See text for the definition of the non-Normality parameter $\theta$.}\label{fig:nonnormal-bias-gamma}
\end{figure*}

\begin{figure*}
\centering\includegraphics{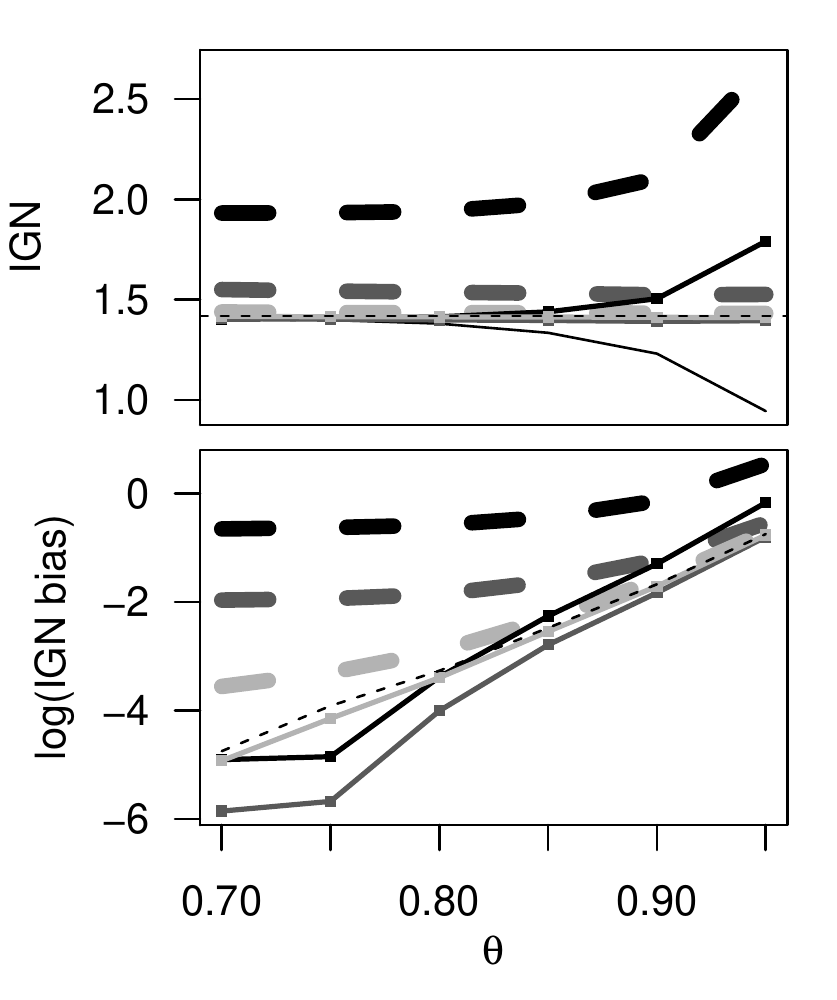}
\caption{Same as \reffig{fig:nonnormal-bias-t} but for bimodal distributions. See text for the definition of the non-Normality parameter $\theta$.}\label{fig:nonnormal-bias-bimodal}
\end{figure*}

We illustrate the effect of nonnormality on the Ignorance score for Normal
ensembles by simulating artificial data from the three types of nonnormal
distributions for different values of the non-Normality parameter $\theta$, and
for different values of the ensemble size $m$. For each combination of $\theta$
and $m$ we have simulated data sets of $10^6$ pairs of ensemble forecasts and
observations. For each data set we have then calculated the average of each of
the four Ignorance scores described above.  The results are illustrated in
\reffig{fig:nonnormal-bias-t}, \reffig{fig:nonnormal-bias-gamma}, and
\reffig{fig:nonnormal-bias-bimodal} for heavy-tailed, skewed, and bimodal
distributions, respectively.

In the upper panel of \reffig{fig:nonnormal-bias-t} the average Ignorance
scores are shown as a function of $\theta$ for the heavy-tailed t-distribution.
Due to convergence to Normality, the score of the continuous Normal
approximation $\mathcal{I}_n$ and the score of the correct t-distribution $\mathcal{I}_p$ 
converge for $\theta\rightarrow\infty$. That is, the bias due to the Normal
approximation vanishes for large $\theta$, as expected. The standard Ignorance
scores of small ensembles are much larger than both, $\mathcal{I}_p$ and $\mathcal{I}_n$. The
differences decrease as the distribution becomes more Normal (i.e.\ for higher
$\theta$), and as the ensembles get larger.  The average values of $\mathcal{I}^*$ are
always closer to the population Ignorance score $\mathcal{I}_p$ of the true
t-distribution than the corresponding values of $\widehat{\mathcal{I}}$, i.e.\ the bias is
reduced by $\mathcal{I}^*$, albeit not removed completely.

The biases of $\mathcal{I}_n$, $\widehat{\mathcal{I}}$ and $\mathcal{I}^*$, i.e.\ their average absolute
differences to the score of the true t-distribution, $\mathcal{I}_p$, are shown in the
lower panel of \reffig{fig:nonnormal-bias-t}. The bias of $\mathcal{I}^*$ is close
to the bias of $\mathcal{I}_n$. That is, the bias correction of $\mathcal{I}^*$ reduces the
bias due the finiteness of the ensemble, and provides an approximation of the
score that an infinitely large ensemble would achieve under a Normal
approximation. The bias of $\mathcal{I}^*$ is consistently smaller than the bias of
$\widehat{\mathcal{I}}$, even though the assumptions that were made to derive $\mathcal{I}^*$ are
not satisfied.  In summary, $\mathcal{I}^*$ cannot remove the bias due to the deviation
from Normality, but it does reduce the bias due to the finiteness of the
ensemble.

\reffig{fig:nonnormal-bias-gamma}, which summarizes the results for the
skewed ensemble data, looks qualitatively similar to 
\reffig{fig:nonnormal-bias-t}. One striking difference is that, for large
ensembles with $m=50$ members, the bias of $\mathcal{I}^*$ is considerably larger than
the bias of $\mathcal{I}_n$ in the case of the t-distribution (\reffig{fig:nonnormal-bias-t}), but it is almost identical to the bias of $\mathcal{I}_n$
for the Gamma distribution. That means that $\mathcal{I}^*$ provides a better
approximation to $\mathcal{I}_n$ in the skewed case than in the heavy-tailed case.  The
biases in the bimodal case are illustrated in 
\reffig{fig:nonnormal-bias-bimodal}. As for the other two types of non-Normality,
the biases of $\mathcal{I}^*$ are generally smaller than the biases of $\widehat{\mathcal{I}}$.
Interestingly, the bias of $\mathcal{I}^*$ is even smaller than the bias of $\mathcal{I}_n$ in
some cases.  The general message from \reffig{fig:nonnormal-bias-t},
\reffig{fig:nonnormal-bias-gamma}, and \reffig{fig:nonnormal-bias-bimodal} is clear:
$\mathcal{I}^*$ is systematically less biased than $\widehat{\mathcal{I}}$, even if the ensemble
members are not exactly Normal distributed.

\end{appendix}

\bibliographystyle{abbrvnat}
\bibliography{debias-ign-articleclass}

\end{document}